 \documentclass[final,3p,twocolumn]{elsarticle}

\usepackage{amssymb}

\journal{Physica B}

\begin{document}

\begin{frontmatter}




\title{100 MHz high-speed strain monitor using fiber Bragg grating and optical filter applied for magnetostriction measurements of cobaltite at magnetic fields beyond 100 T}



\author[lab1]{Akihiko Ikeda\corref{cor1}}
\ead{ikeda@issp.u-tokyo.ac.jp}
\author[lab1]{Toshihiro Nomura}
\author[lab1]{Yasuhiro H. Matsuda\corref{cor1}}
\ead{ymatsuda@issp.u-tokyo.ac.jp}
\author[lab2]{Shuntaro Tani}
\author[lab2]{Yohei Kobayashi}
\author[lab3]{Hiroshi Watanabe}
\author[lab4]{Keisuke~Sato}

\cortext[cor1]{Corresponding author}

\address[lab1]{International MegaGauss Science Laboratory (IMGSL), Institute for Solid State Physics, University of Tokyo, Chiba, Japan}
\address[lab2]{Laser and Synchrotron Research Center (LASOR), Institute for Solid State Physics, University of Tokyo, Chiba, Japan}
\address[lab3]{Graduate School of Frontier Bioscience, Osaka University, Osaka, Japan}
\address[lab4]{Department of Natural Science, Ibaraki National College of Technology, Ibaraki, Japan}

\begin{abstract}
High-speed 100 MHz strain monitor using fiber Bragg grating (FBG) and an optical filter has been devised for the magnetostriction measurements under ultrahigh magnetic fields.
The longitudinal magnetostriction of LaCoO$_{3}$ has been measured at room temperature, 115, 7 and 4.2 K up to the maximum magnetic field of 150 T.
The field-induced lattice elongations are observed, which are attributed to the spin-state crossover from the low-spin ground state to excited spin-states.

\end{abstract}

\begin{keyword}
Fiber Bragg grating, Magnetostriction, Spin crossover, Cobaltite, Ultrahigh magnetic fields



\end{keyword}

\end{frontmatter}


\section{Introduction}
Magnetoelastic response such as magnetostriction is a sensitive indicator of the spin-lattice coupling in materials. 
Magnetostriction is widely observed in systems such as ferromagnets, antiferromagnets \cite{NomuraPRL2014}, spin crossover system \cite{Moaz, ai_lco, ai_pycco, Rotter}, valence transition system, Heavy fermion system \cite{TerashimaJPSJ}, correlated transition metal oxides, multiferroic system, and quantum spin systems \cite{Jaime_PNAS2012}.

In the spin-state crossover systems, volume changes are expected to accompany the spin-state change due to the large difference of the ionic radius of high-spin state and low-spin state.
Recently, we report unusual magnetic field-temperature ($B$-$T$) phase diagram of LaCoO$_{3}$, which is revealed by magnetization measurement up to 133 T \cite{ai_lco}.
The phase diagram is theoretically attributed to the the filed-induced the excitonic insulator phase and the spin-state crystallized phase \cite{Tatsuno, Sotnikov}, which occur due to the spontaneous breaking of the Gauge symmetry and translational symmetry, respectively.
To test this possibility experimentally, direct information on spin-state evolution as a function of magnetic field and temperature is helpful.
Volume change can complement the magnetization data in the sense volume change is directly related to the spin-state change whereas magnetization is affected by both spin-state occupancy and the magnetic interactions.

In this paper, we report a novel system for magnetostriction measurements at $B>100$ T. Using the developed system, we measured the magnetostriction of LaCoO$_{3}$ at magnetic fields beyond 100 T at several temperatures.

\begin{figure*}
\begin{center}
\includegraphics[scale=0.6, clip]{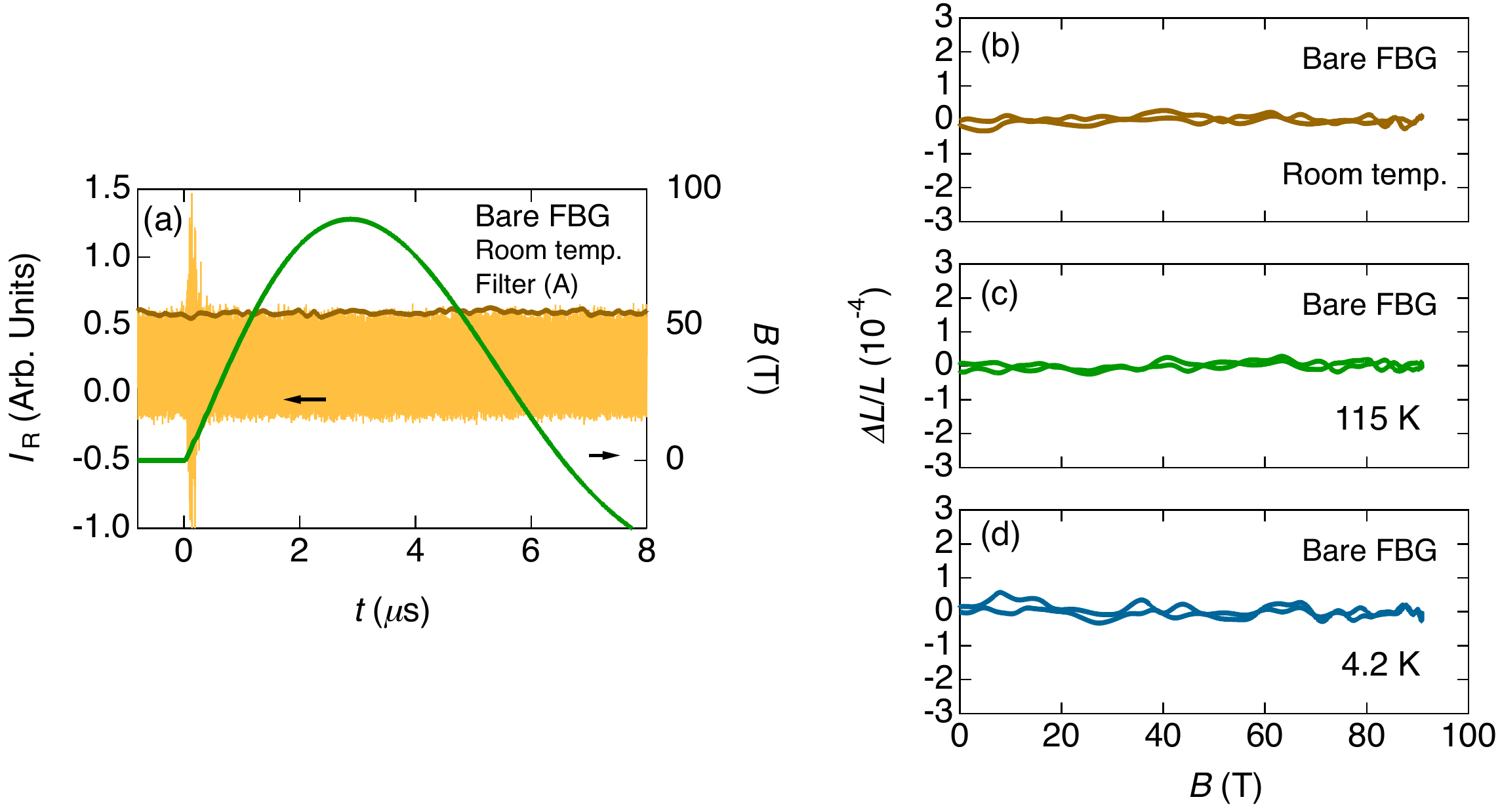} 
\caption{(a) Time evolution of the intensity of reflected light by FBG $I_{\rm{R}}$ and magnetic field at room temperature with a bare FBG suspended in the pulsed magnetic field. Magnetic field dependence of the strain $\Delta L/L$ of bare FBG suspended in the pulsed magnetic fields at (b) room temperature, (c) 115 K and (d) 4.2 K. The reflection from the FBG does not show a noticeable  dependence on the magnetic field. \label{blank}}
\end{center}
\end{figure*}

%

\section{Experiment}
The magnetostriction measurement at above 100 T has been carried out with the combination of the ultrahigh magnetic field pulse and the newly devised high-speed 100 MHz strain monitor based on FBG and a tunable optical filter \cite{ai_fbg}.
The pulsed ultrahigh magnetic fields were generated with the single turn coil (STC) system in the Institute for Solid State Physics, the University of Tokyo, Japan.
Due to the limited pulse duration of STC, the magnetostriction measurement needs to be performed at least 10 MHz repetition rate.
We have developed a high-speed 100 MHz strain monitor using FBG and an optical filter \cite{ai_fbg}.
The longitudinal magnetostriction of the sample is measured through the red and blue shift of the FBG peak.
For the fast detection of the FBG peak position, it is converted to the intensity of the reflected light by passing through an optical filter whose transmission abruptly changes at the relevant wavelength range.
Details of the developed high-speed strain monitor are presented in Ref. \cite{ai_fbg}. A bare glass fiber with FBG was attached to the specimen of a single crystallin LaCoO$_{3}$ in parallel with [110] direction. Optical filter (B) was used in the measurement \cite {ai_fbg}.

Prior to the magnetostriction measurement of LaCoO$_{3}$, we have performed the strain measurements of a bare FBG under pulsed magnetic fields at room temperature, 115 K and 4.2 K.
This is to verify whether there is any magnetic field effects on the FBG reflection at high magnetic fields even without samples.
The results of the measurements are presented in Figs. \ref{blank}(a) to \ref{blank}(d), where the optical filter (A) was used.
As can be seen in Fig. \ref{blank}(a), there is no noticeable variation of the intensity of the FBG reflection $I_{\rm{R}}$ during the pulsed magnetic field is generated.
The magnetic field dependence of the strain $\Delta L/L$ of the bare FBG deduced of measured $I_{\rm{R}}$ at room temperature, 115 K and 4.2 K are presented in Figs. \ref{blank}(b), \ref{blank}(c) and \ref{blank}(d), respectively.
There are no noticeable variation of $\Delta L/L$ at each temperatures within the present experimental error, which is a few 10$^{-5}$.
Thus, it is concluded that there are no intrinsic magnetic field effects on FBG in the present measurements.

\begin{figure*}
\begin{center}
\includegraphics[scale=0.55, clip]{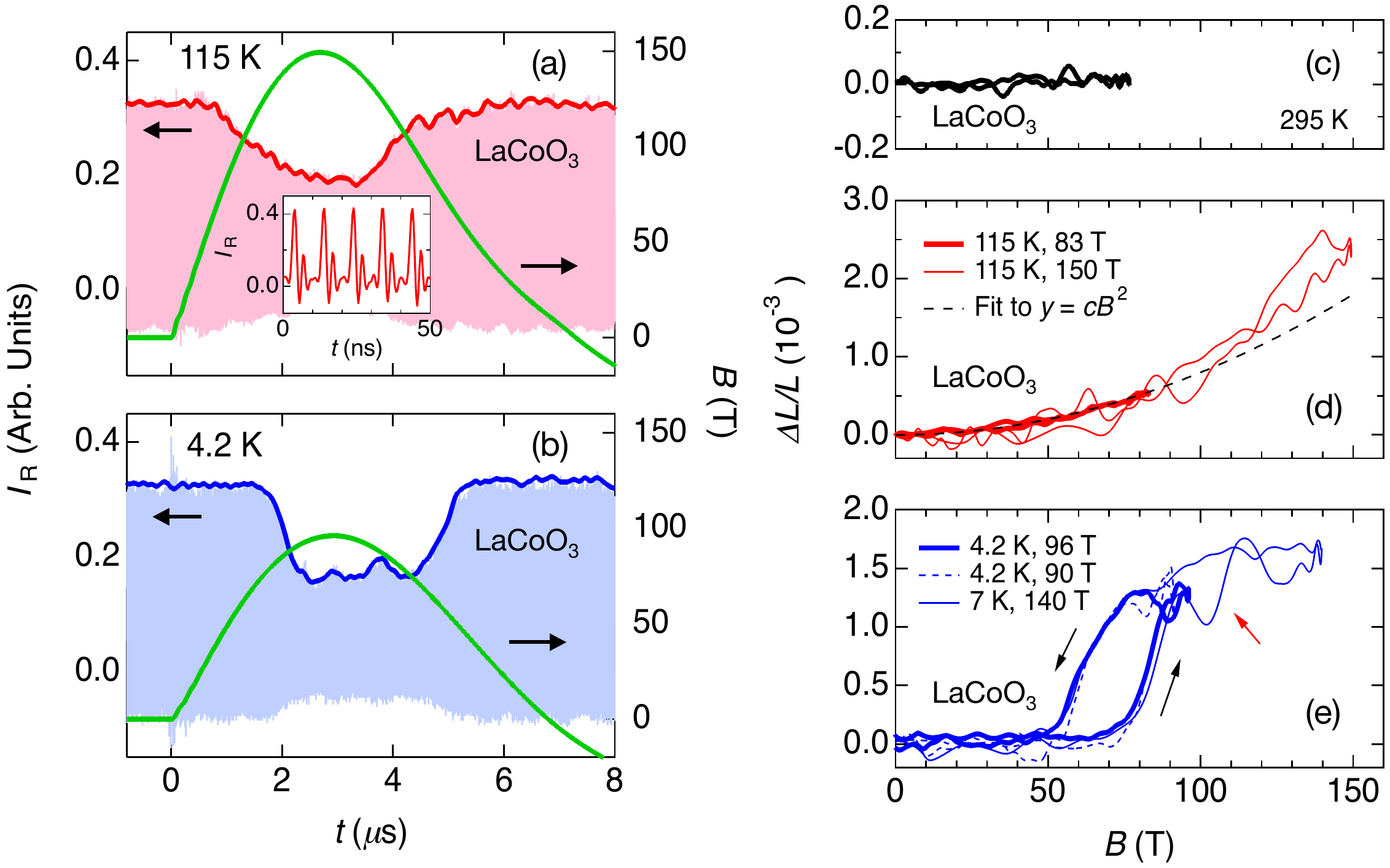} 
\caption{Time evolution of the intensity of reflected light by FBG glued to LaCoO$_{3}$ and magnetic fields at (a) 115 K and (b) 4.2 K. The magnetostriction of LaCoO$_{3}$ as a function of magnetic field at (c) room temperature (d) 115 K and (e) 7 and 4.2 K \label{result}}
\end{center}
\end{figure*}

\section{Result and discussion}
Intensity of light reflected by the FBG attached to a specimen of a single crystalline LaCoO$_{3}$ and the magnetic field were successfully obtained as a function of time as shown in Fig. \ref{result}(a) and \ref{result}(b).
The data are converted to the magnetostriction shown in Figs. \ref{result}(c), \ref{result}(d) and \ref{result}(e).

At room temperature, the magnetostriction is negligibly small within the present experimental error as shown in Figs. \ref{result}(c).
This observation is in good agreement with the previous magnetostriction study of LaCoO$_{3}$ up to 45 T \cite{SatoPhD}.
At room temperature, on the other hand, magnetization is sufficiently large at 33 T \cite{Hoch}.
These facts indicate that the field-induced spin state evolution is negligible at room temperature up to 80 T and that the magnetization increases with magnetic fields mainly due to the paramagnetic susceptibility.
Above observation indicates that magnetostriction is more directly sensing the spin-state evolution as compared to magnetization.

At 115 K, the magnetostriction data shows $B^{2}$ dependence as shown in Fig. \ref{result}(d).
The fit to $y=cB^{2}$ shows a good agreement below 100 T and the deviation is seen in the field range  100 - 150 T.
Whereas, the magnetization shows a linear dependence on the magnetic field up to 33 T \cite{Hoch}.
The $B^{2}$ dependences of magnetostriction and the linear $B$ dependence of magnetization are in good agreement with the previous observation and up to 45 T and model calculation \cite{Sato2008}.
These magnetic field dependences of magnetostriction and magnetization are deduced in models in the limit where the spin-state changes are small and far from the HS saturation.
Thus, the observed result indicates that the system is far from saturation to HS state even at 150 T.

At 7 and 4.2 K, the first order transition of magnetostriction is observed with a large hysteresis.
This is in good agreement with the previous magnetization \cite{Moaz, Sato2009, ai_lco} and magnetostriction measurements \cite{Moaz, Rotter}.
The observed first order transition of magnetostriction shows similar magnetic field dependence with the magnetization \cite{ai_lco}, unlike the results at 115 K and room temperatures.
This fact indicates, magnetization is not disturbed by thermal effects at 7 and 4.2 K, resulting in the observation that magnetization follows the result of the magnetostriction, that is, the spin-state change.

\section{Summary}

The longitudinal magnetostriction of LaCoO$_{3}$ has been measured at room temperature, 115 K and 4.2 K up to the maximum field of 150 T, using the devised high-speed 100 MHz strain monitor.
The field-induced lattice elongations are observed, which are attributed to the spin-state crossover from the LS ground state to excited spin-states.
The data are discussed by comparing the magnetostriction data and the magnetization data.
The magnetostriction data is plausible in the sense that it is directly related to the spin-state change.

\section{acknowledgments}
This work was supported by JSPS KAKENHI Grant-in-Aid for Young Scientists (B) Grant No. 16K17738, Grant-in-Aid for Scientific Research (B) Grant No. 16H04009 and the internal research grant from ISSP, UTokyo.



\bibliographystyle{elsarticle-num} 
\bibliography{fbg}


\end{document}